%% file: main.tex
\documentclass[11pt]{article}
\usepackage[utf8]{inputenc}
\usepackage{amsmath}
\usepackage{graphicx}
\usepackage{svg}
\usepackage{float}
\usepackage{geometry}
\usepackage{subcaption}
\usepackage[numbers, sort, comma, square]{natbib}
\usepackage{booktabs}

\title{Weak Stefan Formulation for Bulk Crystal Growth with Non-smooth Interfaces}
\author{Eyan P. Noronha, B. Erik Ydstie}
\date{\today}

\begin{document}

\maketitle
\begin{abstract}
    Most heat transfer models for bulk crystal growth rely on the classical Stefan formulation to evaluate interface motion during phase change. However, when the interface is non-smooth the use of the classical Stefan formulation may lead to singularities. To address this problem, we propose a simulation model based on the weak formulation of the Stefan problem. Numerical solutions of the weak Stefan formulation are obtained using the finite volume method. This approach provides an energy conserving discretization scheme that accurately evaluates heat transfer around non-smooth interfaces. We apply the weak formulation to numerically simulate the solidification of silicon in the horizontal ribbon growth process. Results exhibit a limitation on the ribbon's pull speed, which previous classical Stefan models have failed to demonstrate. A comparison of heat transfer between radiation and gas cooling shows that gas cooling increases the pull speed limit for the same amount of heat removed.

\end{abstract}

\input{sections/introduction.tex}
\input{sections/formulation.tex}
\input{sections/numerical.tex}
\input{sections/results.tex}

\input{sections/conclusions.tex}

\bibliography{main}
\bibliographystyle{unsrtnat}

\end{document}

%% file: sections/introduction.tex
\section{Introduction}
Simulation and modeling play an important role in the safe and reliable operation of crystal growth furnaces \cite{derby2018synergy}. At their core, crystal growth models describe a process of phase transition taking place across a moving interface. The physics of phase transition can be complicated, and will depend on the scale of the process and the properties of the materials involved. Still, a first order model for phase transition can be developed, based on the basic principles of mass and energy conservation. These models belong to a well known class of mathematical formulations known as the Stefan problem~\cite{visintin2008introduction}. 

Solving a Stefan problem in 2 or 3 dimensions can be challenging since the shape of the interface is not known beforehand. Therefore, crystal growth models often rely on numerical techniques to obtain accurate solutions~\cite{lan2004recent}. Methods based on Galerkin finite element combined with arbitary Lagrangian-Eulerian (ALE) approach have been used in a variety of crystal growth models due to their high spatial accuracy~\cite{ghosh1993arbitrary,zhang2020analysis}. \citet{helenbrook2018high} developed an adaptive mesh algorithm to track the interface using triangular, ALE moving mesh to tackle large unsteady interfacial deformations. Finite volume methods using interface tracking algorithms are used in \cite{fainberg2007new,jung2013combined} to build 3D models for the Czochralski process. \citet{weinstein2019modeling} applied a Lattice Boltzmann model to track interface evolution while accounting for anisotropic interface attachment kinetics. Other approaches that track the interface evolution implicitly, like phase field models, are also used to model crystal growth~\cite{yan2019formation}. However, they are better suited to model phase transition at the microscopic level based on thermodynamic considerations, like dendrite growth and phase boundaries~\cite{jhang2019three,jokisaari2018phase}.

Despite the ubiquity of Stefan problem in modeling crystal growth processes, certain problems exist in its application. One major problem is the use of the strong formulation, also known as the classical formulation, of the Stefan problem to model non-smooth interfaces. An interface is termed non-smooth if the unit normal at any point on the interface is not uniquely defined. On a smooth interface the strong form of the Stefan condition based on energy conservation is given by
\begin{equation}\label{eq:classical_stefan2}
    \Big[k_l\nabla T_l-k_s\nabla T_s\Big]\cdot\,\hat{n}=-\rho L \,\Vec{v}\cdot\hat{n},
\end{equation}
where $\hat{n}$ is a unit vector normal to the interface; $k_s, k_l$ are the thermal conductivities and $\nabla T_s, \nabla T_l$ are the temperature gradients for the solid and liquid phase. $\rho$ denotes the density of the solid phase, $L$ is the latent heat of fusion and $\Vec{v}\cdot\hat{n}$ denotes the velocity of the interface from solid to liquid. Equation~\eqref{eq:classical_stefan2} holds point-wise everywhere on the interface.

\begin{figure}[t]
\centering
  \centering
  \includegraphics[width=\linewidth]{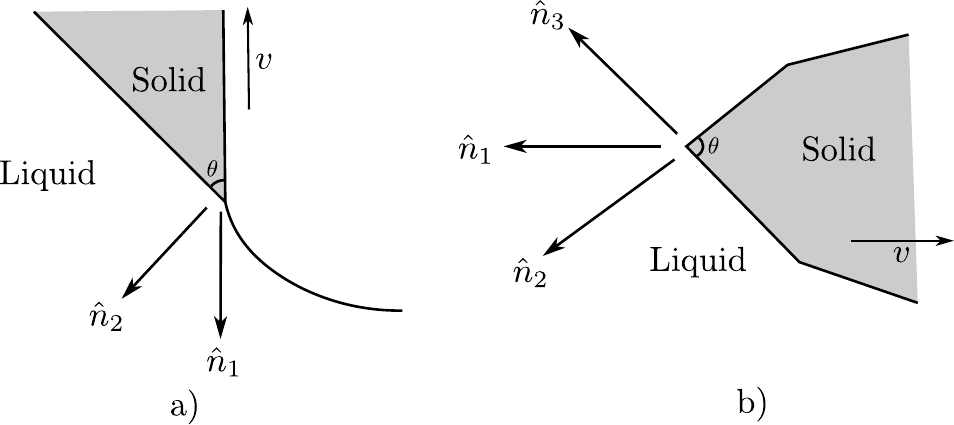}
  \caption{Examples of ill-defined interface velocity at non-smooth corners in bulk crystal growth}
  \label{fig:motivation}
\end{figure}

The assumption on the smoothness of the interface puts a restriction on the applicability of the classical Stefan formulation. This is especially relevant for crystal growth systems since the occurrence of non-smooth interfaces is routine~\cite{krauze20193d,yan2019formation,schwabe2020spiral}. Figure~\ref{fig:motivation} provides two example of frequently encountered situations in crystal growth where the classical Stefan formulation is not applicable. Figure~\ref{fig:motivation}a is a close-up of a crystal growth configuration around a triple point. The solidification interface makes an angle $\theta$ with the free surface. In the case of a faceted growth, $\theta$ can be a fixed angle~\cite{stockmeier2018edge,stockmeier2019analysis, kellerman2016floating}. The unit normal is uniquely defined everywhere on the interface except at the triple point. At the triple point, the tip grows with a velocity of either $v$, if measured from the direction of $\hat{n}_1$;  or $v \sin\theta$, if measured from $\hat{n}_2$. This presents an ambiguity in the choice of the normal velocity in the Stefan condition \eqref{eq:classical_stefan2}. A similar problem exists in systems with facet growth~\cite{fujiwara2012crystal,lau2020situ,hu2020situ}, as shown in figure~\ref{fig:motivation}b, where the normal component of the interfacial velocity cannot be uniquely defined.

Using the classical Stefan formulation to model interfaces that are non-smooth can lead to incorrect results~\cite{segal1998conserving,vuik2000conserving,king1999two}. This is because the limit of the energy balance equation on an arbitrarily small neighbourhood of a non-smooth interface does not exist and only the integral form of the energy balance makes sense. In the context of bulk crystal growth systems, the use of the classical formulation has created singularities around triple points and facet corners, which has made the analysis of heat transfer difficult.

In the Bridgman process, theoretical and numerical solutions have shown singularities at the intersection of crucible walls with the interface~\cite{yekel2017,kuiken1988note}. This has lead to the phenomenon of ``Interface effect", which makes it difficult to control the interface shape~\cite{ostrogorsky2018interface,volz2009interface}. Singularities have also been observed near the triple point in the Czochralski process~\cite{sackinger1989finite,ramachandran2009comparison}. Asymptotic analysis have found the order of these singularities to be dependant on the angle prescribed by the triple point or the facet corners~\cite{anderson1994fluid,wigley1969method}. As a result, the use of numerical techniques that rely on piecewise polynomials may not represent heat transfer at such points~\cite{cheng2014analysis}.  

\begin{figure}[hb]
\centering
  \centering
  \includegraphics[width=1\linewidth]{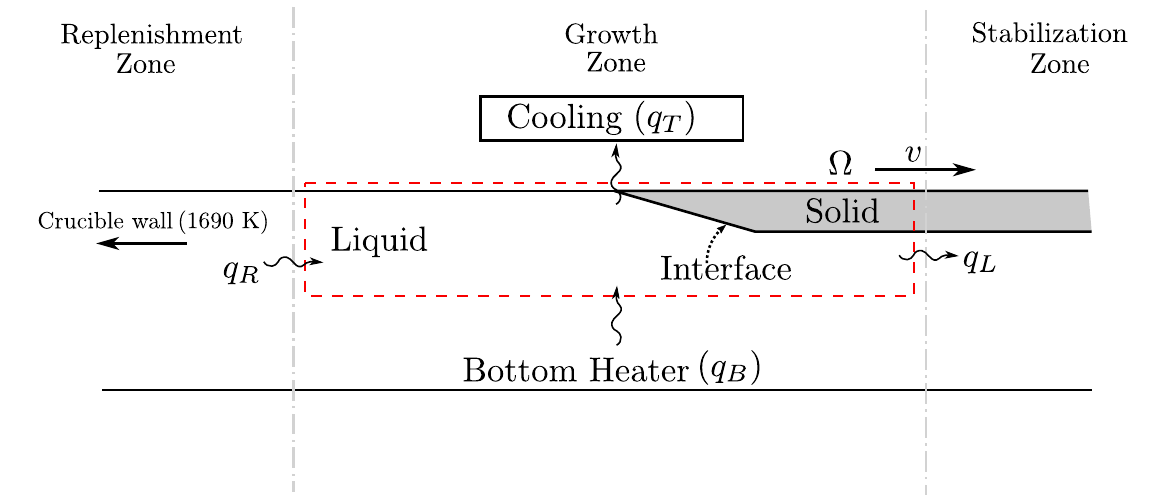}
  \caption{Schematic of the horizontal ribbon growth process around the growth zone}
  \label{fig:Schematic}
\end{figure}

One particular crystal growth method where singularities have made it difficult to provide robust predictions is the horizontal ribbon growth (HRG) process~\cite{glicksman1983analysis,helenbrook2015solidification,pirnia2020analysis}. In this process, thin sheets of single crystal silicon are produced directly from the melt. Figure~\ref{fig:Schematic} provides a schematic of the growth zone in the horizontal ribbon growth furnace. Cooling is applied at the top surface of a molten bath and a seed crystal is inserted horizontally to nucleate the growth process~\cite{oliveros2015existence}. The thickness of the growing crystal is controlled by manipulating the cooling rate and the pull speed $(v)$. A higher pull speed provides less exposure to cooling and thus produces thinner ribbons. However, experiments have shown that the HRG process exhibits a limitation on its pull speed, which also constraints the size of the ribbon's thickness~\cite{kellerman2013floating}. Theoretical modeling and simulations have not provided any conclusive explanation for the pull speed limit~\cite{zoutendyk1978theoretical,zoutendyk1980analysis,daggolu2012thermal,daggolu2013stability,daggolu2014analysis, helenbrook2016experimental}. This has hindered the scale up of the process, since it has not been possible to increase production speed to meet industrial standards~\cite{greenlee2015towards}.

The aim of this paper is to develop a theoretical framework for simulating crystal growth based on the weak formulation of the Stefan problem. We apply this theory to the numerical simulation of the horizontal ribbon growth process. At the same time, we keep the formulation general enough to be applicable to other areas of crystal growth as well. Although fluid flow, meniscus stability and impurities play an important role in the stable operation of the horizontal ribbon growth process \cite{sun2018simulating,sun2020numerical,ke2017effects,noronha2020weierstrass}, here we decouple these phenomena and focus only on the conductive heat transfer aspect of the process. In doing so, we are able to provide an explanation for the occurrence of some phenomenon, like the limitation on pull speed, observed in experiments. 

%% file: sections/formulation.tex
\section{Mathematical Formulation}

Consider the process of solidification to take place inside a bounded domain $\Omega$, belonging to a subset of the Euclidean space $\mathbf{R}^n$ $(n=2,3)$. We shall assume that the process occurs at a constant volume, i.e., no volume change associated with phase change. The domain is occupied by a pure material capable of attaining two phases, liquid and solid, at a sharp melting point $T_m$. To simplify fluid motion, the material is assumed to move at a uniform speed, $v$, throughout the domain. The analysis is carried out in a Lagrangian frame of reference, so the material appears stationary in this reference frame. 
 
 The time interval for analysis is set to $]0,\tau [$, where $\tau$ is a constant and define $Q:=\Omega\times]0,\tau [$. An energy balance carried out over any volume $V$ with surface area $A$ gives
\begin{equation}\label{eq:weak_form1}
    \int_\Gamma \Bigg[ \frac{\partial}{\partial t} \int_V U dV + \oint_A \Vec{f}\cdot\hat{n}\, dA \Bigg] dt =0 \qquad \forall \quad V\times\Gamma \subseteq Q
\end{equation}
where $U$ is the density of internal energy within volume $V$ and $\Vec{f}\cdot\hat{n}$ is the flux of the energy transferred through the boundary $A$. The energy balance holds over any time interval $\Gamma \subseteq ]0,\tau[$. The use of the integral form is advantageous as it reduces the requirements on the regularity of the solution, for in general $U$ and $\Vec{f}$ will be discontinuous at the phase interfaces. Since we carry out our analysis in a Lagrangian reference frame, conduction is the only main source of energy transport within the material. Therefore, the energy flux at any point inside the domain is given by the Fourier law:
\begin{equation}
    \Vec{f}=-k\nabla T.
    \label{eq:flux}
\end{equation}
Additional information on the nature of $U$ will be needed for \eqref{eq:weak_form1} to be suitable in practise. This will be given by an energy-phase rule, which is a modification of the temperature-phase rule in \cite{visintin2008introduction}, to account for metastable states.

\subsection{The Energy-Phase Rule} 
\begin{figure}[ht]
\centering
  \centering
  \includegraphics[width=0.6\linewidth]{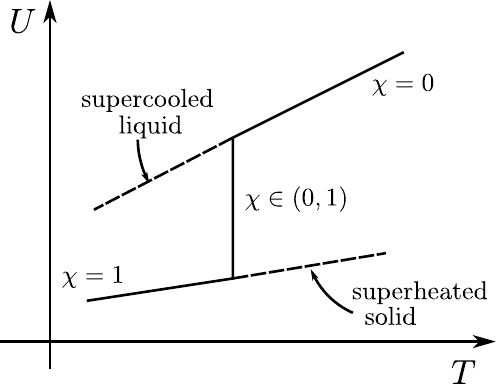}
  \caption{Energy-Phase rule}
  \label{fig:EPrule}
\end{figure}

We define a solid fraction field, $\chi\in[0,1]$, to denote the phase of the material at any point in $Q$. $\chi=1$ corresponds to the material in the solid phase and $\chi=0$ to the material in the liquid phase. An intermediate value of $\chi$ may denote a fine solid-liquid mixture at the macroscopic length scales. For our formulation, we require that the phase change only occurs at the melting point $T_m$, during which the temperature remains constant until $\chi$ changes phase to either 0 or 1. The formulation should also allow for the existence of supercooling in absence of nucleation during extreme cooling. These features are captured through an energy-phase rule of the form given below:
\begin{align}
    U=
    \begin{cases}
    \rho c_{s}(T-T_m) & \chi=1\\
    \rho L \chi &\chi\in(0,1),\, T=T_m\\
    \rho L+\rho c_{l}(T-T_m) & \chi=0.
    \end{cases}
    \label{eq:energy_phase}
\end{align}

Figure~\ref{fig:EPrule} illustrates the essential features of the energy-phase rule. The internal energy is considered to be a multi-valued function of the temperature that branches at two points. The choice of the branch depends on whether a phase change is initiated as the temperature approaches the melting point. The criteria for initiating a phase change is determined by a set of rules discussed in section 4.1. Since metastable states, like a supercooled melt, are often observed in practise, this energy-phase rule allows us to model the physics of crystal growth in a manner that is representative of experimental systems. 

Substituting \eqref{eq:energy_phase} in \eqref{eq:weak_form1}, the energy balance equation can be re-written as:
\begin{equation}\label{eq:weak_form2}
    \int_\Gamma \Bigg[ \frac{\partial}{\partial t} \int_V (\rho c_\chi (T-T_m)+\rho L \chi) dV + \oint_A \Vec{f}.\hat{n} dA \Bigg] dt =0 \qquad \forall \quad V\times\Gamma \subseteq Q,
\end{equation}
where $c_\chi$ is equal to $c_s$ in the solid phase and equals to $c_l$ in the liquid phase. 

%% file: sections/numerical.tex
\section{Simulation Modeling}
\begin{figure}[ht]
\centering
  \centering
  \includegraphics[width=1\linewidth]{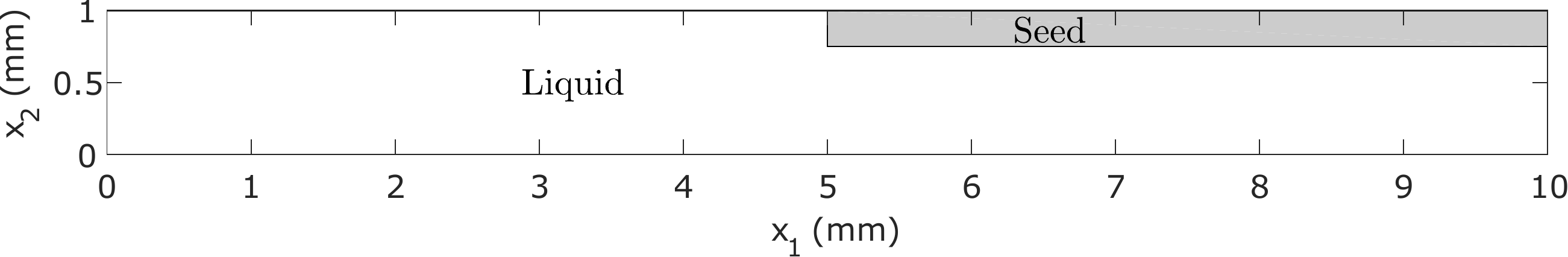}
  \caption{Simulation domain and initial seed configuration of the HRG process}
  \label{fig:initial}
\end{figure}

To test the utility of our weak formulation, we carry out finite volume simulations to model the growth of a silicon ribbon in a horizontal ribbon growth process. The simulation is carried out in a rectangular domain of size $1$mm$\times10$mm around the growth tip of the ribbon. The shape and size of control volumes are usually chosen based on the geometry of the domain and the accuracy needed for the required engineering application. For our simulations, we divide the domain into rectangular control volumes $(V_i)$ of size $\Delta x=5\mu$m. The time interval for simulation will be divided into equal open intervals $\Gamma_n$ of size $\Delta t$. 

Each cell element $V_i\times\Gamma_n$ of $Q$ will be categorized to either of the three domains: solid $Q_s$, liquid $Q_l$ or interface $Q_I$. Phase change will be restricted to take place only in the interface domain $Q_I$. In the solid $Q_s$ or liquid $Q_l$ domain, no phase change occurs so the energy balance equation \eqref{eq:weak_form2} reads
\begin{equation}\label{eq:temp_update}
    \int_{\Gamma_n}\Bigg[\rho c_{j}\frac{\partial \overline{T_i}}{\partial t}\Delta x-Df_i\Bigg] dt=0\qquad \forall\quad V_i\times\Gamma_n\subseteq Q_{j}, \quad j=\{s,l\} ,
\end{equation}
where $\overline{T_i}$ is the temperature field averaged over the control volume $V_i$. $D$ is the difference operator and $Df_i=f_R-f_L+f_T-f_B$ denotes the intensity of the heat flux removed from the control volume $V_i$. $f_R,f_L,f_T$ and $f_B$ represent the heat fluxes from the right, left, top and bottom walls of $V_i$ respectively.

The interface is characterized as a region of phase change with constant melting point $T_m$. Therefore, the energy balance equation \eqref{eq:weak_form2} in the interface domain $(Q_I)$ is written as
\begin{align}
    \int_{\Gamma_n}\Bigg[\rho L \frac{\partial \overline{\chi_i}}{\partial t}\Delta x-Df_i\Bigg]dt=0 \qquad \forall \quad V_i\times\Gamma_n\subseteq Q_I
    \label{eq:phase_update}
\end{align}
where $\overline{\chi_i}$ is the solid fraction averaged over the control volume $V_i$. Equation~\eqref{eq:phase_update} equates the intensity of latent heat released due to phase change to the intensity of heat removed from a control volume of size $\Delta x$. 

Equation~\eqref{eq:temp_update} and \eqref{eq:phase_update} are conservative discretizations of the balance equation \eqref{eq:weak_form2}, i.e., the flux on the boundary of one cell equals the flux on the boundary of the adjacent cell. Since the conservation holds at the discrete level; if the numerical method converges, they can be proven to converge to a weak solution of the conservation law \eqref{eq:weak_form2} using the Lax-Wendroff theorem~\cite{lax1960systems}.

In a Stefan problem, the solid-liquid interface behaves like a moving boundary that evolves continuously over time. Due to this, the proposed simulation scheme proposed relies on the categorization of each cell element at the beginning of every time step. Therefore, a capture rule is required to determine the propagation of the interface for numerical simulations. 

\subsection{Interface Propagation}

In the classical Stefan formulation, phase transition occurs only at the interface, between the boundaries of the solid and liquid phases. This condition serves as a useful abstraction to model solidification processes involving metastable states, namely, supercooling and superheating. In contrast, the enthalpy based methods for solidification assume thermodynamic equilibrium at all points in the domain.

Keeping this in mind, we constraint local thermodynamic equilibrium to occur only along a small band of size $\Delta x$ between the solid and the liquid phase. Physically, this would signify the proximity of the liquid to nucleation sites for phase transition. This region of size $\Delta x$ will be defined as the interface. Away from the interface, we allow for the possibility of metastable states. All of this is governed by the energy phase rule described in \eqref{eq:energy_phase}. 

\begin{figure}[ht]
\centering
  \centering
  \includegraphics[width=0.9\linewidth]{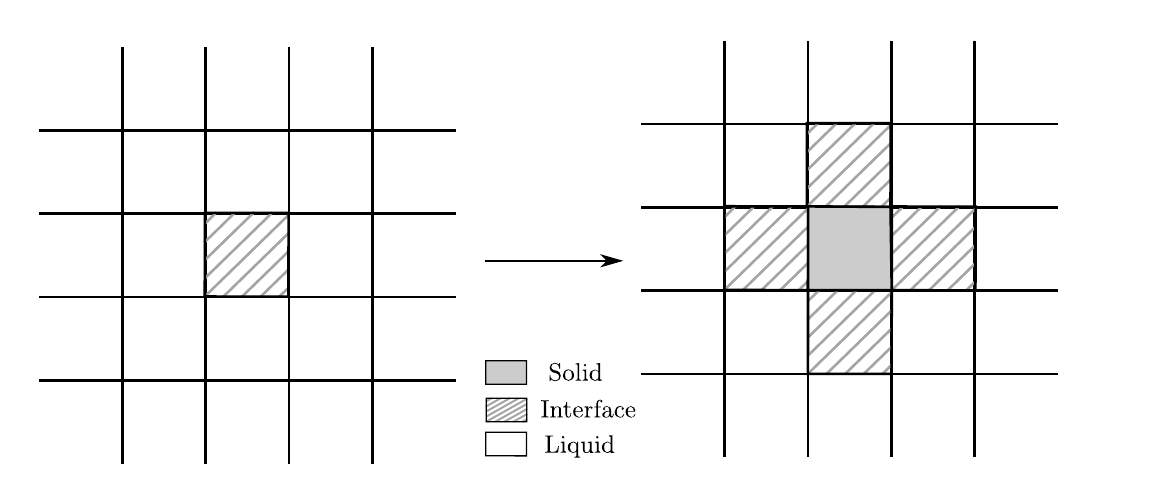}
  \caption{Interface propagation rule for solidification.}
  \label{fig:front}
\end{figure}

Initially, each cell in $\Omega\times\{0\}$ is categorized as either a solid, liquid or interface. At the end of each time interval, an interface propagation step is carried out. In this step, any cell belonging to the interface that exceeds its solid fraction beyond $[0,1]$ changes its label to either liquid or solid appropriately. This transformed cell acts as a nucleation site for its nearby cells to become a part of the interface. Figure~\ref{fig:front} shows an example of a $\Delta x$ radial neighbourhood, also called the Von Neumann neighbourhood, used for interface propagation. In general, crystals have anisotropic surface energy, so the interface propagation may be preferred along certain directions. However, in this paper we consider the case when the crystal growth is isotropic.   

The interface propagation step is carried out at the end points of the discretization scheme. Due to this, additional steps need to be taken to ensure energy is conserved when the labels on the cells are changed. When a cell transforms from a solid or liquid phase to an interface, the temperature of the cell is updated to $T_m$ and the residual thermal energy is transferred into the solid fraction using
\begin{equation}\label{eq:residual}
    L\,\Delta\overline{\chi}_i=c_\chi (T_m-\overline{T}_i),
\end{equation}
where $\Delta\overline{\chi}_i$ is the change in the solid fraction of cell $i$. A similar strategy is applied when the solid fraction of an interface cell exceeds $[0,1]$ and changes to either a solid or a liquid cell. 

\subsection{Numerical Method}

Applying \eqref{eq:flux} to each wall in the control volume $V_i$, the flux can be discretized using a central difference scheme and substituted in \eqref{eq:temp_update} and \eqref{eq:phase_update}. Combined with the boundary conditions, the system of equations describing the evolution of the temperature and phase fields is complete and can be solved using any standard method of numerical integration. For this paper, we found the Douglas-Gunn Alternating Direction Implicit (ADI) scheme to be stable and efficient in solving the integral equations. The implementation of the ADI scheme is standard and we refer the readers to \citet{mcdonough2008lectures} for more information. 

Due to the Lagrangian nature of the simulation, the domain needs to be re-centered after a fixed number of iterations to prevent the domain from leaving the area of interest. To do this effectively, we re-center the simulation domain after every $\Delta x/v\Delta t$ iterations. This restrict the choice of $v$ so that $\Delta x/v\Delta t$ is an integer. The simulation is said to reach steady-state when the solution becomes time independent within an acceptable level of tolerance. 

At the initial stages of the simulation, the numerical integration starts with large step sizes, $\Delta t \approx 10 \Delta x$. The step size is then gradually decreased to $\Delta t=\Delta x$, until steady state is reached. In some situations, it was observed that the interface would oscillate for a large number of iterations and would not reach steady state. This was attributed to the explicit nature of the interface propagation step. The conversion of residual solid fraction into thermal energy based on \eqref{eq:residual} could sometimes cause the algorithm to not converge due to the large value of latent heat $L$. In this case, further decrease in the step size of the simulation or distribution of the excess solid fraction to the nearby transformed cells was found to be sufficient to stop the oscillations. 

%% file: sections/results.tex
\newpage
\section{Application: Horizontal Ribbon Growth}
To determine the boundary conditions required for our simulations, we refer to Figure \ref{fig:Schematic} for a cross-sectional schematic of the horizontal ribbon growth furnace. A cooling mechanism at the top surface of the melt drives the solidification process which causes single crystal silicon to grow. Cooling takes place either through a passive mechanism, like radiation or through an active cooling system, like gas jets~\cite{kudo1980improvements,cheng1993growth,jewett1981method,helenbrook2015solidification}. It is unclear how one cooling mechanism may provide an advantage over the other. Therefore, we consider both mechanisms of heat removal at the top boundary condition in our simulation study.

For the case of radiation, the top surface heat loss is given by the Stefan-Boltzmann law
\begin{equation}
    q_t(x_1)=\epsilon\sigma(T^4-T_c^4)F(x_1),
\end{equation}
where $\sigma$ is the Stefan-Boltzmann constant, $\epsilon=\chi\epsilon_s+(1-\chi)\epsilon_l$ is the weighted emmisivity of solid $(\epsilon_s)$ and liquid $(\epsilon_l)$ emmisivities. $T_c=300$K is the temperature of the water cooled walls of the furnace surrounding the crucible. $F(x_1)$ is called the view-factor and takes into account the area exposed by an opening, like a slit, to the water cooled walls of the furnace from any point $x_1$ on the surface of the melt. We consider the slit width to be variable and placed above the top surface of the melt at a height $h$. The width of the slit is parameterized using the variable $w$. The view-factor at any point $x_1$ on the surface of the melt is given by the formula~\cite{howel2011thermal}:

\begin{equation}
    F(x)=\frac{\sin \phi_2-\sin \phi_1}{2},
\end{equation}
where $\sin \phi_1$ and $\sin\phi_2$ are
\begin{equation}
    \sin\phi_1=\frac{-(w+x)}{\sqrt{(w+x)^2+h^2}}\qquad\sin\phi_2=\frac{(w-x)}{\sqrt{(w-x)^2+h^2}}.
\end{equation}
For the base case, a value of $w=5$mm is chosen. A small value of $h$, say $h=0.1$mm, allows a smooth transition in the radiative heat flux near the slit edges. The variable $w$ will be used as a parameter to study the effects of radiation length on pull speed.

To model the gas cooling jet, a scaled down version of the experimental conditions in \citet{helenbrook2016experimental} will be used. For gas cooling (hereon referred as Gaussian cooling), the top surface heat removal rate is modeled using a Gaussian curve, parameterized by peak intensity $q_{peak}$ and spread $\sigma$. 
\begin{equation}
    q_t(x)=q_{peak}\;\exp{\Big(\frac{-x^2}{2\sigma^2}\Big)}.
\end{equation}

An approximate fit to the experimental data in \citet{helenbrook2016experimental} was found at $\sigma=0.8$. $q_{peak}$ was chosen to be $40$ $W/cm^2$, to ensure the heat removed using the Gaussian cooling profile was of the same order of magnitude as the heat removed using radiation. These values of $\sigma$ and $q_{peak}$ will serve as the base case for Gaussian cooling and provide us with a comparative study of the two cooling mechanisms.

A positive thermal gradient in the melt is found to provide a stable environment for crystal growth~\cite{fujiwara2012crystal}. For this reason, a small heat flux is applied on the bottom boundary of the domain, supplied from the heaters on the underside of the crucible. 
\begin{equation}
    q_B=2\;\; W/cm^2
\end{equation}

\begin{figure}[ht]
\centering
  \centering
  \includegraphics[width=1\linewidth]{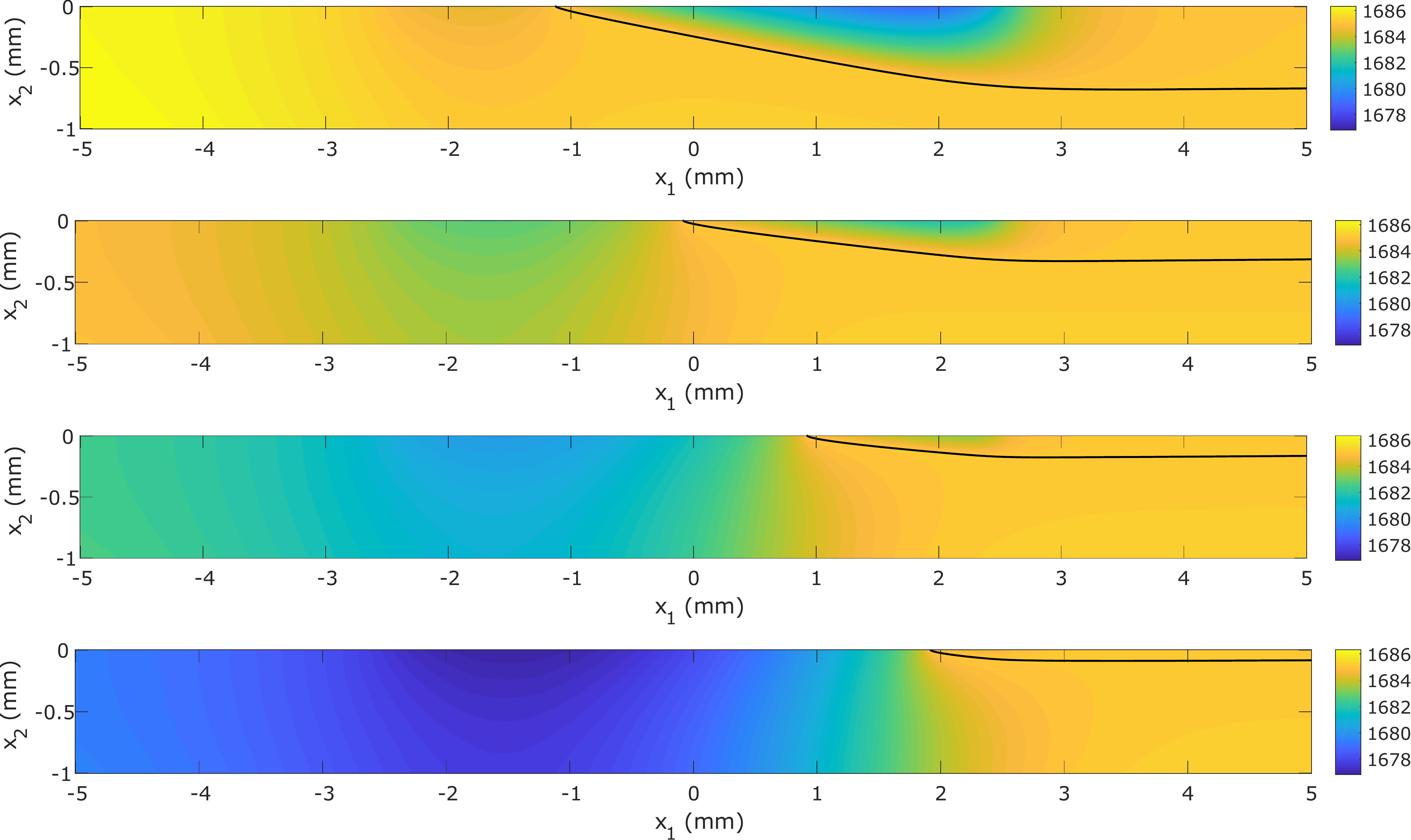}
  \caption{Temperature maps for radiation cooling at ribbon pull speeds of 0.3, 0.5, 0.7, 0.9 mm/sec from top to bottom, respectively. The solid black line describes the position of the solid-liquid interface.}
  \label{fig:radiation_results_2}
\end{figure}

In a reference frame moving at a constant velocity $v$, the top and bottom boundary conditions appear to drift in the opposite direction. To account for this, we introduce a time dependent drift on the top boundary condition.
\begin{equation*}
    q_T(x_1,t)=q_t(x_1-vt).
\end{equation*}
Since the bottom flux is a constant, it remains unchanged.

Preceding the growth zone is a replenishment zone that provides a constant supply of melt. This is done by means of heaters inside the crucible walls that melt silicon feed chunks. Experimental conditions maintain the temperature of the silicon melt at $1690$K~\cite{helenbrook2016experimental}. This bulk melt at $1690$K is assumed to be $5cm$ away from the simulation domain. We also assume a linear temperature profile inside the replenishment zone with respect to the moving reference frame. The left boundary condition is therefore calculated to be
\begin{equation}
    q_L=-k_l\frac{\Delta T}{\Delta x}=\frac{67\times 10^-4}{5\times10^{-2}}(1690-T)=0.134\times(1690-T)\;\;W/cm^2
\end{equation}

As the ribbon is pulled out of the simulation domain, it exits into a stabilization zone. In this region, the temperature in the solid and liquid phases are maintained so that conduction only occurs in the vertical direction~\cite{helenbrook2016experimental}. Although convective heat transport may still exist in the horizontal direction, in a moving reference frame this effect is not realized and therefore the heat flux at the right boundary will be taken to be zero.
\begin{equation}
    q_R=0.
\end{equation}

The initial condition for simulations was found to be fairly robust to any appropriate choice of seed crystal shape and temperature field. The first simulation was initialized using a rectangular seed crystal of width $250\mu m$ and length $5$mm, covering the top half surface of the melt as shown in figure~\ref{fig:initial}. The melt was initialized with a uniform temperature of $1690K$ and the solid was chosen to be at a $1680K$. Subsequent simulation for different operating condition were initialized from the steady state solution of the previous simulation.

\subsection{Results}
\begin{table}
\begin{center}
\begin{tabular}{ccc}
\toprule
Parameter	&Symbol   &Value\\[3pt]
\midrule
Density of liquid silicon		&$\rho$		&2530 $[\text{kg/m}^3]$\\
Thermal conductivity of silicon melt			&$k_l$			&67  $[\text{W/mK}]$\\
Thermal conductivity of silicon solid		&$k_s$			&22 $[\text{W/mK}]$\\
Heat capacity of silicon melt	&$c_l$							&1000 [J/kgK]\\
Heat capacity of silicon solid &$c_s$		&1060 [J/kgK]\\
Latent heat of fusion  &$L$		&$1.8\times10^{6}$ [J/kg]\\
Emmisivity of silicon melt	&$\epsilon_l$							&0.2\\
Emmisivity of silicon solid &$\epsilon_s$		&0.6\\
\bottomrule
\end{tabular}
\caption{Material properties of silicon used for simulation}
\label{tab:properties}
\end{center}
\end{table}

We begin the simulation study by applying the finite volume discretization to the base cases of radiation and Gaussian cooling. The values of the physical constants required for simulation are summarized in Table 1.

\begin{figure}[H]
\centering
  \centering
  \includegraphics[width=0.6\linewidth]{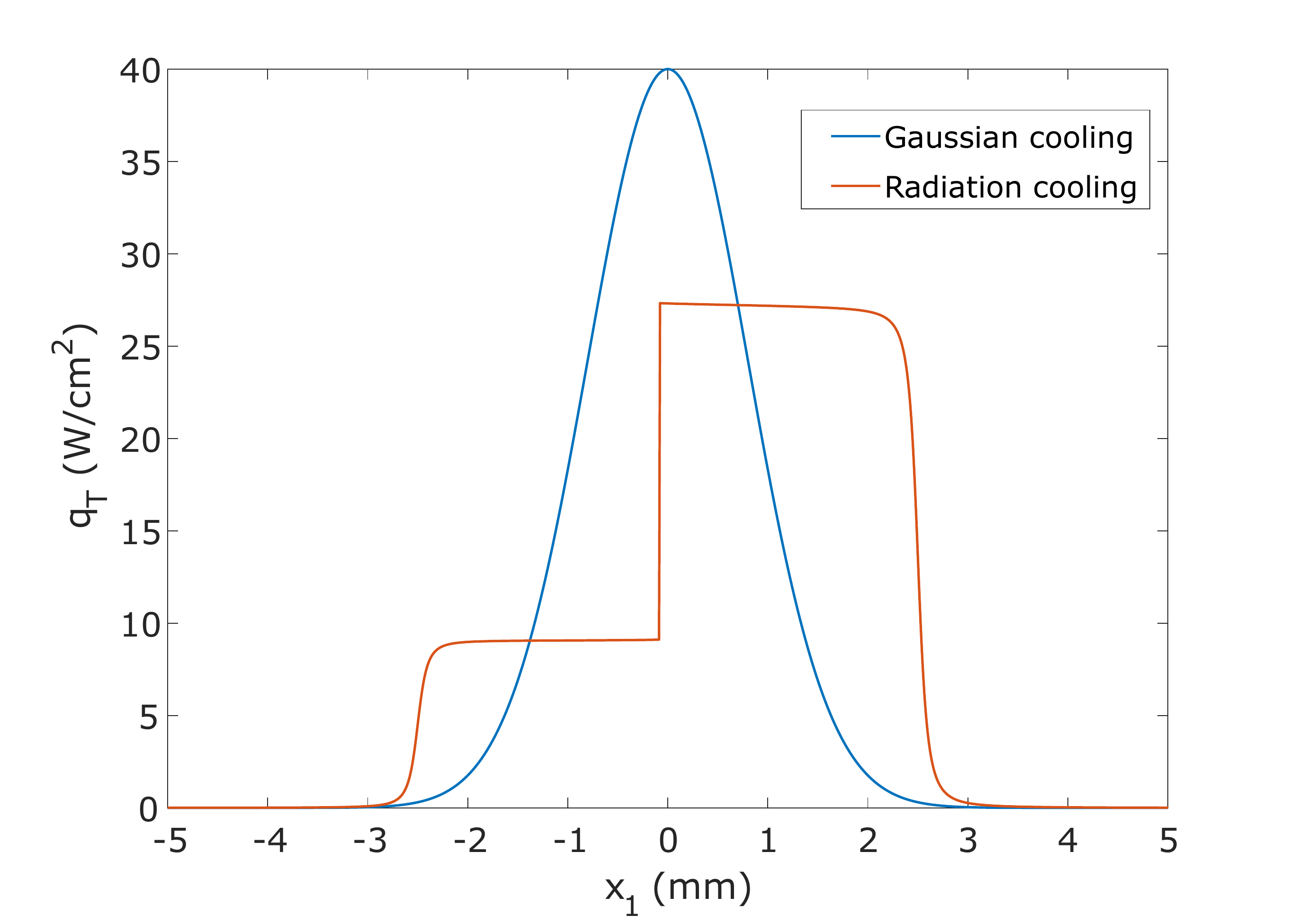}
  \caption{Comparison of top surface heat flux for radiation ($v=0.5$mm/sec) and Gaussian cooling}
  \label{fig:cooling}
\end{figure}

We first consider the base case when the top surface of the melt is cooled by radiation. For this case, the slit only allows the center $5$ mm of the melt to radiate heat, which acts as the sole mechanism of latent heat removal. We perform simulations at pull speed increments of $0.05$ mm/sec, starting from $0.3$ mm/sec. Figure~\ref{fig:radiation_results_2} illustrates the steady-state temperature fields at pull speeds of $v=0.3,0.5,0.7$ and $0.9$ mm/sec. The solid black line in the figures denotes the shape of the ribbon. For all pull speeds, the ribbon shape is found to approximate a wedge shape. This can be attributed to the near constant heat flux at the top surface of the solid \cite{zoutendyk1978theoretical}. Figure~\ref{fig:cooling} illustrates the top surface heat flux for radiation and Gaussian cooling. The radiative heat flux was plotted at a steady-state pull speed of $v=0.5$ mm/sec. The heat flux appears constant in solid and liquid phases and shows a jump in between due to their difference in emmisivities.

\begin{figure}[ht]
\centering
  \centering
  \includegraphics[width=1\linewidth]{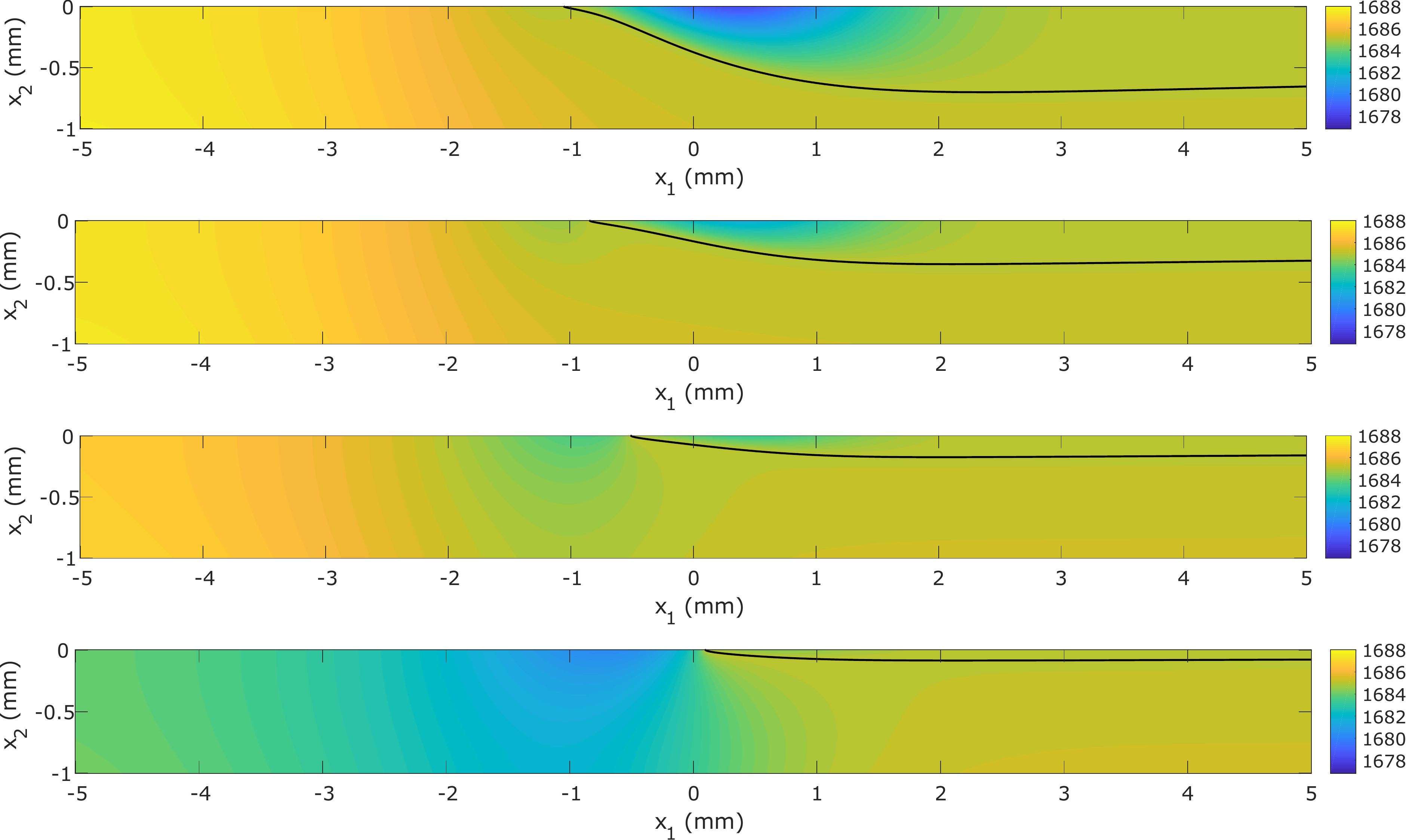}
  \caption{Temperature maps for gas cooling at ribbon pull speeds of 0.2, 0.4, 0.8, 1.6 mm/sec from top to bottom, respectively.}
  \label{fig:gaussian_results_2}
\end{figure}

For the base case involving Gaussian cooling, simulations performed at 0.1 mm/sec increments, starting from a pull speed of $0.2$ mm/sec. Unlike the radiation case, the heat removed by Gaussian cooling does not depend on the position of the ribbon tip. The temperature field and ribbon shape at pull speeds of $v=0.2, 0.4,0.8$ and $1.6$ mm/sec are displayed in Figure~\ref{fig:gaussian_results_2}. In this case, the ribbon shape is curved due to the non-linear shape of the Gaussian cooling profile.  

Comparing figures \ref{fig:radiation_results_2} and \ref{fig:gaussian_results_2}, we observe certain similarities and differences between the two cooling mechanisms. In both cases, we observe the ribbon tip moving to the right as the pull speed increases. This creates a U-shaped pool of supercooled melt in front of the ribbon tip that grows larger in size. It is worthwhile to note the larger pool size in radiation compared to Gaussian cooling. Another important observation is the motion of the ribbon tip to the center in the Gaussian case, and to the edge of the slit in the radiation case. 

\begin{figure}[htb]
\centering
  \centering
  \includegraphics[width=0.65\linewidth]{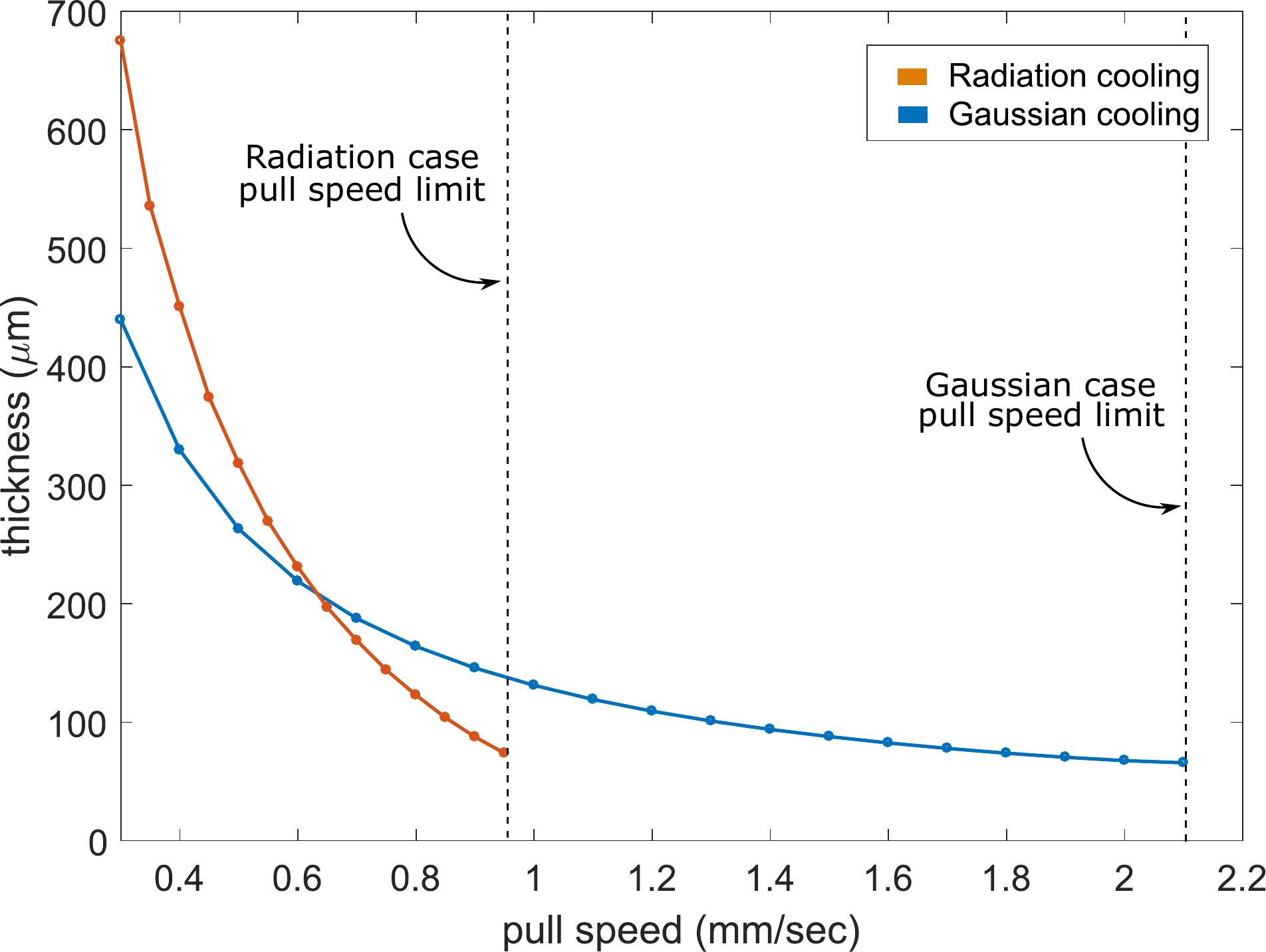}
  \caption{Thickness vs pull speed plot for radiation and Gaussian cooling. The dashed lines signify the pull speed limitation observed in simulations.}
  \label{fig:thickness}
\end{figure}

The information on the pull speed and ribbon thickness for radiation and Gaussian cooling is summarized as a graph in figure~\ref{fig:thickness}. This relationship is governed by the total energy balance equation
\begin{equation}\label{eq:vtbalance}
    L\rho v t_r=Q_{tot},
\end{equation}
where $t_r$ is the thickness of the ribbon and $Q_{tot}$ is the total heat removed from the domain. The two curves intersect at a pull speed of $v=0.63$ mm/sec. Therefore, the point where the two curves intersect denotes equal heat removal $Q_{tot}$ for the two cases. Below this pull speed the heat removed from the radiative case is higher and above this pull speed the heat removed is lower due to the difference in emmisivities of the two phases. This causes radiation to produce thicker ribbons at lower pull speeds and thinner ribbons at higher pull speeds. 

In Addition, we also observe a pull speed limitation for both mechanisms. For radiation, the limit occurs at 0.9 mm/sec as the ribbon tip reaches the end of the slit. On the other hand, the pull speed limit for Gaussian cooling occurs at 2mm/sec, as the ribbon tip reaches the center of the cooling jet.

\subsubsection{Heat transfer at the triple point}

\begin{figure}[ht]
\centering
  \centering
  \includegraphics[width=\linewidth]{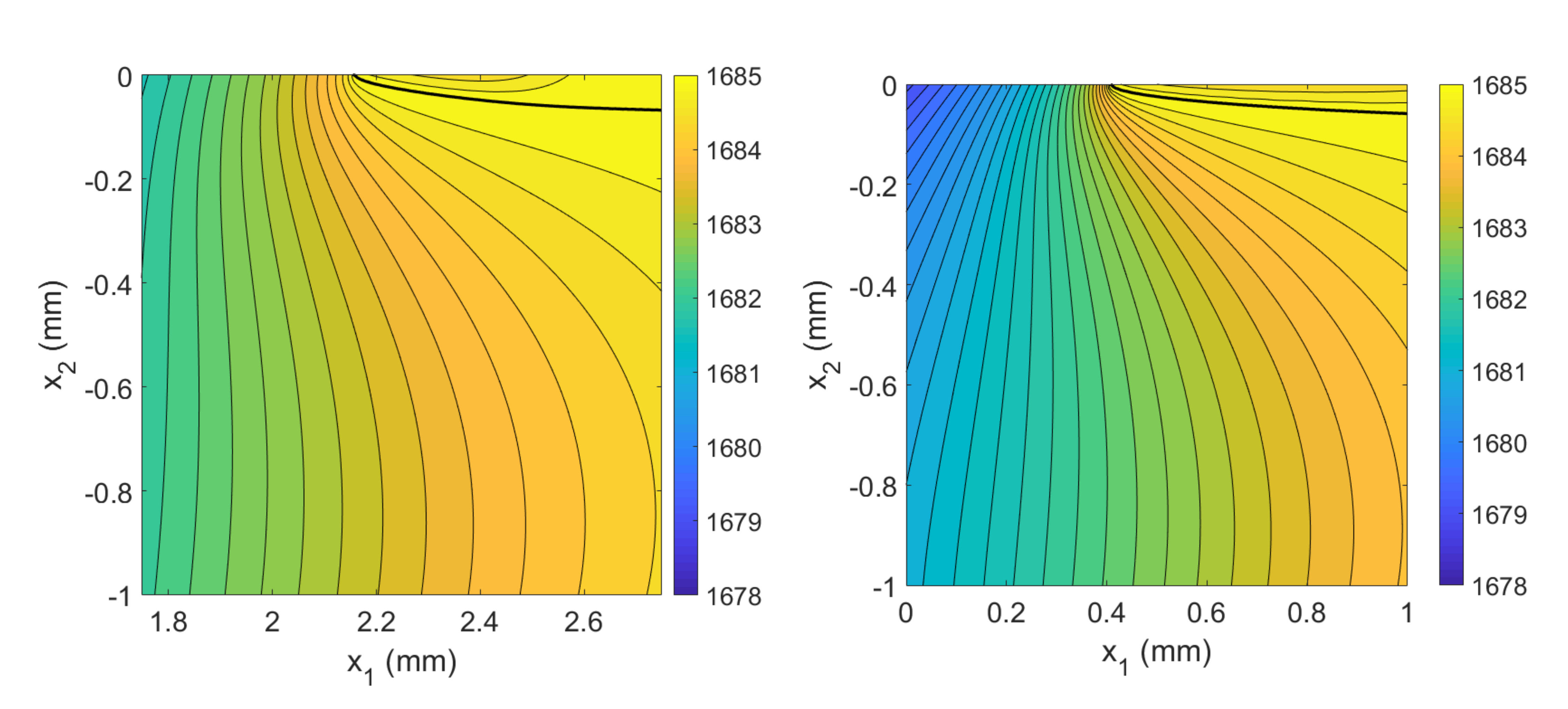}
  \caption{A close up of the temperature field around the ribbon tip. The thick black line denotes the interface at $T_m=1685$K. Thin black lines denote isotherms at 0.2K intervals.}
  \label{fig:Growth_tip}
\end{figure}

To explain the pull speed limitation in our simulations, we take a closer look at the temperature profile near the ribbon tip. Figure~\ref{fig:Growth_tip} shows a close-up of the temperature contours near the ribbon tip for the two base cases at their respective pull speed limit. The isotherms become nearly vertical as they approach the ribbon tip. This implies a predominantly horizontal mode of heat removal at the ribbon tip. This is in contrast to the conventional notion of heat removal in the horizontal ribbon growth process, which was considered to be in the vertical direction. This provides an indication to the origin of pull speed limit due to heat transfer.

\begin{figure}[ht]
\centering
  \centering
  \includegraphics[width=\linewidth]{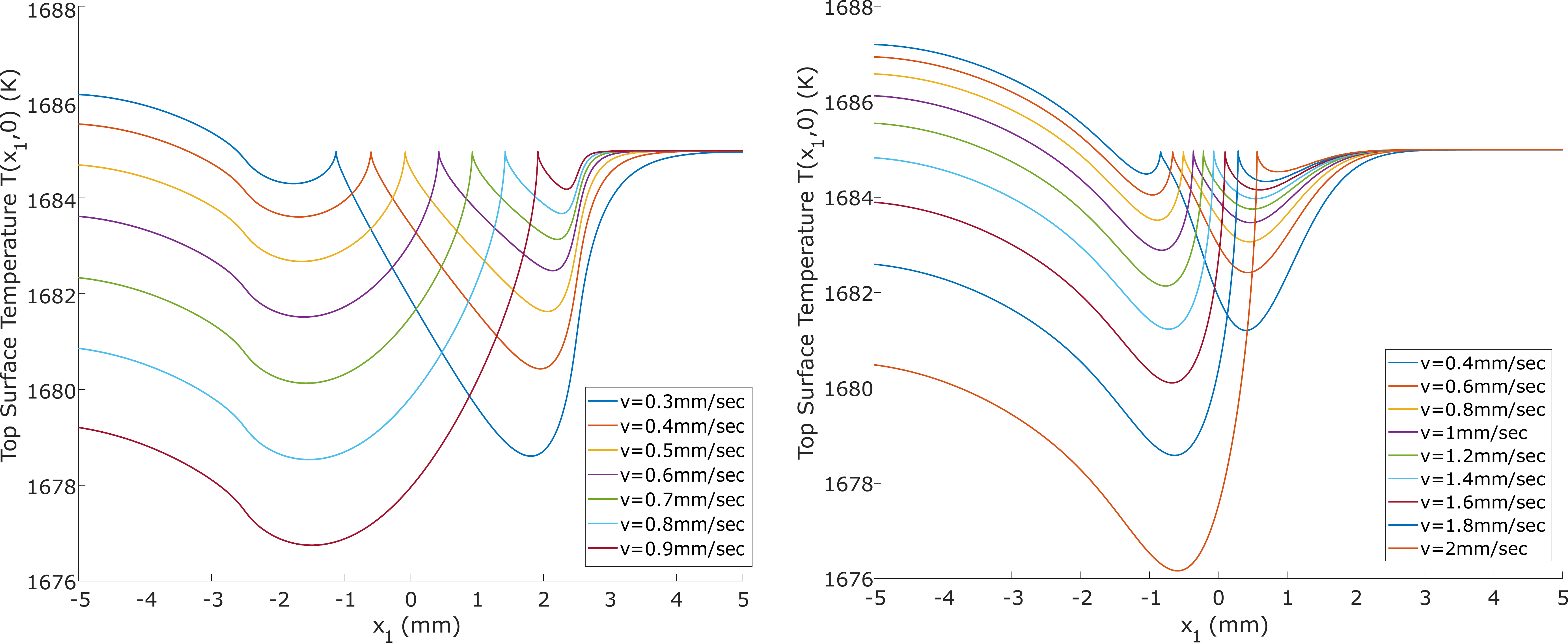}
  \caption{Top surface temperature profile for radiation (left) and Gaussian (right) cooling.}
  \label{fig:top_temp}
\end{figure}

To explain this hypothesis, we study the variations in the temperature gradient around the ribbon tip for increasing pull speeds. Figure \ref{fig:top_temp} plots the temperature profile at the top surface of the domain, $T(x_1,0)$ for the two base cases. The multiple curves in each plot, from left to right, denote the temperature profile for increasing pull speeds. The position of the ribbon tip in these curves can be identified by the peak at $T_m=1685$K. At low pull speeds, almost all the heat removal required to maintain solidification at the ribbon tip is from the solid side. As the pull speed is increased, the ribbon tip shifts to the right. This decreases the heat removed from the solid surface and increases the heat removed form the liquid surface. The increase in the heat removed from the liquid side creates a pool of super-cooled melt in front of the ribbon tip. The negative thermal gradient in the melt provides the necessary source of heat removal to maintain growth at the tip. Moreover, the decrease in the heat removed from the solid side causes the ribbon to get thinner. Therefore, from the perspective of heat transfer, it is more efficient to remove heat from the liquid side than the solid side because some portion of the heat removed from the solid side is used to maintain the ribbon's thickness. 

Based on the above explanation we summarize the observation of pull speed limit as follows: increasing the pull speed of the ribbon requires an equal increase in the growth rate of solid at the ribbon tip. If sufficient latent heat is not removed to maintain this growth rate, the ribbon tip moves to the right. This increases the amount of heat removed from the tip---by increasing the heat removed from the liquid side---and establishes a new equilibrium position. As the pull speed is increased further, at some point the heat removed from tip is maximized and the growth rate reaches its limits. For the Gaussian case this limit occurs around the center of the cooling profile, while for the radiation case this limit occurs at the edge of the slit. 

\subsubsection{Parametric Study}
 
\begin{figure}[hb]
\centering
  \centering
  \includegraphics[width=0.62\linewidth]{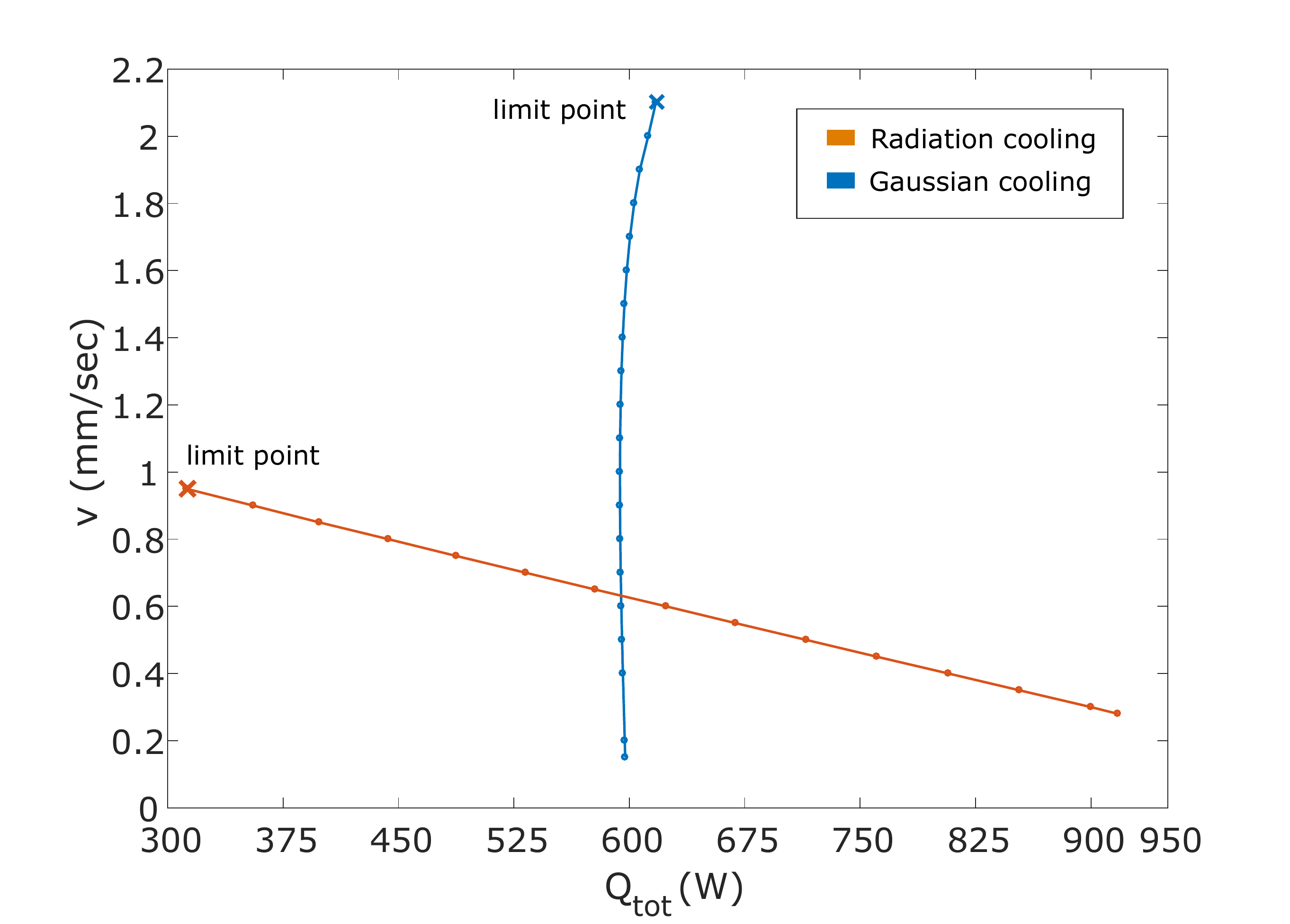}
  \caption{A graph highlighting the variability in total heat removed using radiation as a function of pull speed. In contrast, the heat removed using cooling jet is almost constant.}
  \label{fig:sensitivity1}
\end{figure}

Simulations from the previous section suggest the pull speed limit for Gaussian cooling to be higher than radiation. However, since the heat removed due to radiation varies with pull speed, it is unclear if the advantage lies in the mode of cooling or the quantity of heat removed. Figure \ref{fig:sensitivity1} plots the total amount of heat removed $Q_{tot}$, which includes the sum of conductive and convective heat transport from all four boundaries, at different pull speeds for the two base cases. The heat removed for Gaussian cooling is nearly constant while the heat removed during radiation cooling decreases with increasing pull speed. At the limit point, the heat removed in the radiation case is significantly lower than the Gaussian case. Therefore, for a fair comparison it is reasonable to ask how radiation and Gaussian cooling compare for the same amount of heat removed.   

To do this, a parametric study of the top surface cooling profiles is performed by varying the slit width $w$ in radiation cooling and $q_{peak}$ in Gaussian cooling. For the radiation case, 5 sets of simulations are performed with slit widths of $w=2.5$ mm, 3.75 mm, 5 mm, 6.25 mm, and 7.5 mm. For Gaussian cooling, we choose $q_{peak}=\;$12.48 W/cm$^2$, 18.19 W/cm$^2$, 23.96 W/cm$^2$, 28.94 W/cm$^2$, and 33.91 W/cm$^2$. These values of $q_{peak}$ were chosen to match the total top surface heat removed from the radiation case. 

\begin{figure}[ht]
\centering
  \centering
  \includegraphics[width=0.62\linewidth]{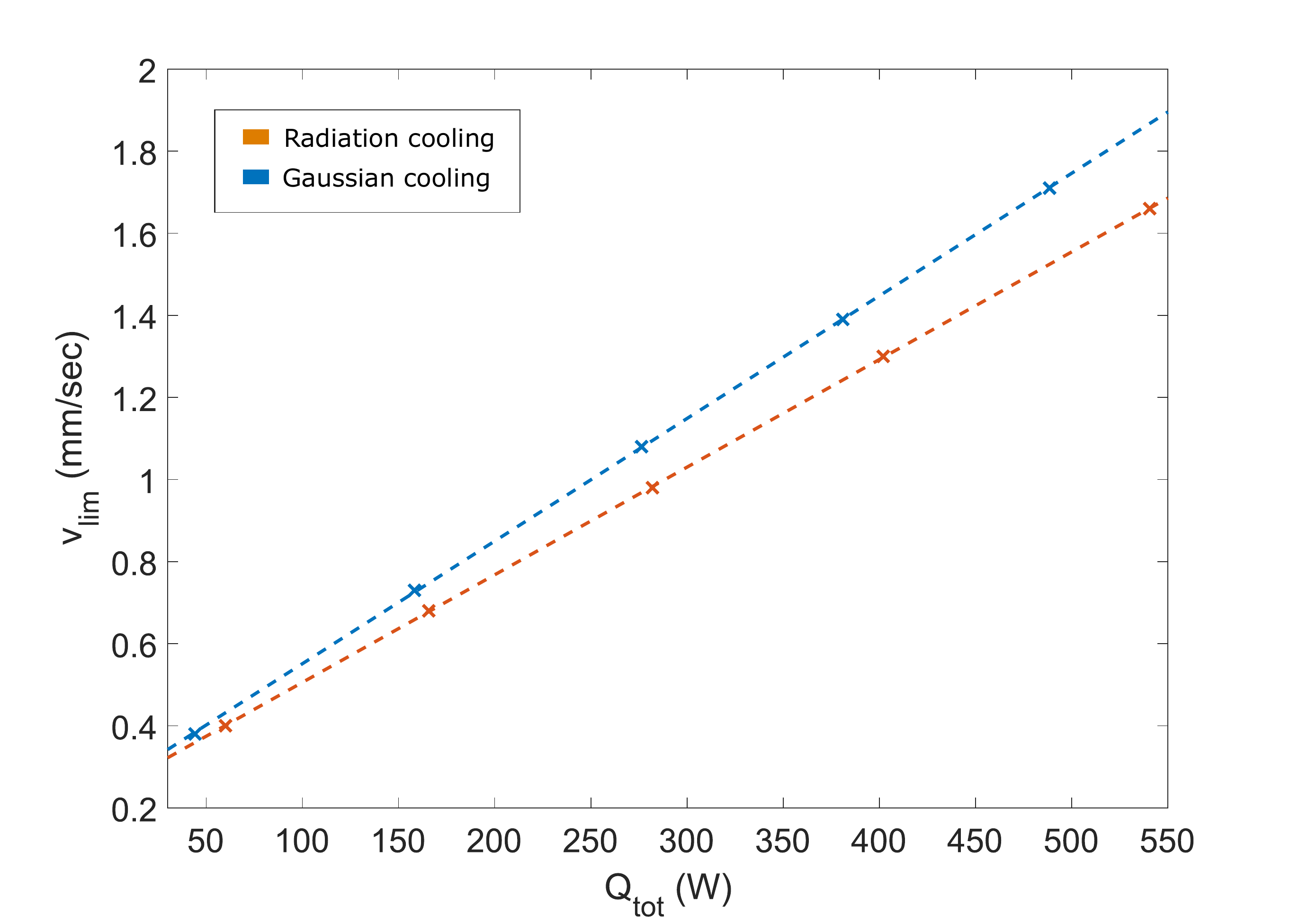}
  \caption{A graph of maximum pull speed as a function of heat removed. The dashed lines denotes the best fit lines.}
  \label{fig:sensitivity2}
\end{figure}

Similar to the process of finding the limit points in figure \ref{fig:sensitivity1}, we find the pull speed limit $v_{lim}$ in each case and mark them in figure~\ref{fig:sensitivity2}. Interestingly, the points follow a straight line with $R^2$ value close to 1 upto 3 decimal places.    

For both mechanisms, increasing the total heat removed from the system $(Q_{tot})$, leads to a proportional increase in the pull speed limit. For a given $(Q_{tot})$, Gaussian cooling achieved a higher maximum pull speed compared to radiation cooling. This can be attributed to the narrow heat removal profile in Gaussian cooling. Therefore, we find that a cooling jet provides better performance in producing high speed ribbons than radiation. The improvement is larger at higher heat removal rates.

%% file: sections/conclusions.tex
\section{Conclusions}
The main goal of this paper is to provide an alternative theoretical framework for simulating crystal growth models involving non-smooth edges. We find that around non-smooth interfaces, like the triple point in a horizontally grown ribbon, the classical models do not satisfy energy conservation. Therefore, a weak formulation of the Stefan problem is needed to relax the requirements on the regularity of the interface. A modified energy-phase rule is derived to account for the existence of metastable states. We choose a finite-volume discretization scheme to maintain the conservative form of the weak formulation. This allows us to come up with an easy to implement simulation scheme, which satisfies energy conservation and provides sufficient accuracy for engineering applications.

As an application, we perform a simulation study of the horizontal ribbon growth process. Unlike the classical Stefan formulations, the weak formulation demonstrates a pull speed limitation as observed in experiments. We explain the pull speed limit based on a local heat transfer arguments. During experiments, this limitation may be compounded by other physical constraints like the formation of a stable meniscus or solidification kinetics \cite{daggolu2013stability,helenbrook2016experimental}. However, from the perspective of heat transfer alone, insufficient heat removal from the growth tip is shown to be the fundamental reason for pull speed limitation.

Two different mechanisms of heat removal are analyzed. We find that a diffuse cooling profile, like radiation, is less effective than a narrow cooling profile, like a cooling jet, in producing thin ribbons at high speed. A linear relationship was discovered between the total heat removed from the furnace and the maximum pull speed in both cooling mechanisms. This relationship provides an interesting insight into the heat transfer occurring in the horizontal ribbon growth furnace and would require further analysis beyond the scope of this simulation. All of this aids in our understanding of the heat transfer conditions required for the high speed operation of the horizontal ribbon growth process.

%% file: main.bbl
\begin{thebibliography}{54}
\providecommand{\natexlab}[1]{#1}
\providecommand{\url}[1]{\texttt{#1}}
\expandafter\ifx\csname urlstyle\endcsname\relax
  \providecommand{\doi}[1]{doi: #1}\else
  \providecommand{\doi}{doi: \begingroup \urlstyle{rm}\Url}\fi

\bibitem[Derby(2018)]{derby2018synergy}
Jeffrey~J Derby.
\newblock The synergy of modeling and novel experiments for melt crystal growth
  research.
\newblock In \emph{IOP Conference Series: Materials Science and Engineering},
  volume 355, page 012001. IOP Publishing, 2018.

\bibitem[Visintin(2008)]{visintin2008introduction}
Augusto Visintin.
\newblock Introduction to stefan-type problems.
\newblock \emph{Handbook of differential equations: evolutionary equations},
  4:\penalty0 377--484, 2008.

\bibitem[Lan(2004)]{lan2004recent}
CW~Lan.
\newblock Recent progress of crystal growth modeling and growth control.
\newblock \emph{Chemical engineering science}, 59\penalty0 (7):\penalty0
  1437--1457, 2004.

\bibitem[Ghosh and Moorthy(1993)]{ghosh1993arbitrary}
Somnath Ghosh and Suresh Moorthy.
\newblock An arbitrary lagrangian-eulerian finite element model for heat
  transfer analysis of solidification processes.
\newblock \emph{Numerical Heat Transfer}, 23\penalty0 (3):\penalty0 327--350,
  1993.

\bibitem[Zhang et~al.(2020)Zhang, Gao, Tremsin, Perrodin, Shalapska, Bourret,
  Onken, Vogel, and Derby]{zhang2020analysis}
Chang Zhang, Bing Gao, Anton~S Tremsin, Didier Perrodin, Tetiana Shalapska,
  Edith~D Bourret, Drew~R Onken, Sven~C Vogel, and Jeffrey~J Derby.
\newblock Analysis of chemical stress and the propensity for cracking during
  the vertical bridgman growth of babrcl: Eu.
\newblock \emph{Journal of Crystal Growth}, 546:\penalty0 125794, 2020.

\bibitem[Helenbrook and Hrdina(2018)]{helenbrook2018high}
BT~Helenbrook and J~Hrdina.
\newblock High-order adaptive arbitrary-lagrangian--eulerian (ale) simulations
  of solidification.
\newblock \emph{Computers \& Fluids}, 167:\penalty0 40--50, 2018.

\bibitem[Fainberg et~al.(2007)Fainberg, Vizman, Friedrich, and
  Mueller]{fainberg2007new}
J~Fainberg, D~Vizman, J~Friedrich, and G~Mueller.
\newblock A new hybrid method for the global modeling of convection in cz
  crystal growth configurations.
\newblock \emph{Journal of crystal growth}, 303\penalty0 (1):\penalty0
  124--134, 2007.

\bibitem[Jung et~al.(2013)Jung, Seebeck, and Friedrich]{jung2013combined}
T~Jung, J~Seebeck, and J~Friedrich.
\newblock Combined global 2d--local 3d modeling of the industrial czochralski
  silicon crystal growth process.
\newblock \emph{Journal of crystal growth}, 368:\penalty0 72--80, 2013.

\bibitem[Weinstein et~al.(2019)Weinstein, Virozub, Miller, and
  Brandon]{weinstein2019modeling}
Oleg Weinstein, Alexander Virozub, Wolfram Miller, and Simon Brandon.
\newblock Modeling anisotropic shape evolution during czochralski growth of
  oxide single crystals.
\newblock \emph{Journal of Crystal Growth}, 509:\penalty0 71--86, 2019.

\bibitem[Yan et~al.(2019)Yan, Wang, Guo, Zhang, Wang, An, and
  Liu]{yan2019formation}
Jing-Yuan Yan, Yong-Wei Wang, Yong-Ming Guo, Wei Zhang, Cong Wang, Bao-Li An,
  and Dong-Fang Liu.
\newblock Formation and preferred growth behavior of grooved seed silicon
  substrate for kerfless technology.
\newblock \emph{Chinese Physics B}, 28\penalty0 (6):\penalty0 066802, 2019.

\bibitem[Jhang and Lan(2019)]{jhang2019three}
JW~Jhang and CW~Lan.
\newblock Three-dimensional phase field modelling of twin nucleation during
  directional solidification of multi-crystalline silicon.
\newblock \emph{Journal of Crystal Growth}, 520:\penalty0 33--41, 2019.

\bibitem[Jokisaari et~al.(2018)Jokisaari, Voorhees, Guyer, Warren, and
  Heinonen]{jokisaari2018phase}
Andrea~M Jokisaari, Peter~W Voorhees, Jonathan~E Guyer, James~A Warren, and
  Olle~G Heinonen.
\newblock Phase field benchmark problems for dendritic growth and linear
  elasticity.
\newblock \emph{Computational Materials Science}, 149:\penalty0 336--347, 2018.

\bibitem[Krauze et~al.(2019)Krauze, Virbulis, Zitzelsberger, and
  Ratnieks]{krauze20193d}
A~Krauze, J~Virbulis, S~Zitzelsberger, and G~Ratnieks.
\newblock 3d modeling of growth ridge and edge facet formation in< 100>
  floating zone silicon crystal growth process.
\newblock \emph{Journal of Crystal Growth}, 520:\penalty0 68--71, 2019.

\bibitem[Schwabe(2020)]{schwabe2020spiral}
Dietrich~G Schwabe.
\newblock Spiral crystal growth in the czochralski process—revisited, with
  new interpretations.
\newblock \emph{Crystal Research and Technology}, 55\penalty0 (2):\penalty0
  1900073, 2020.

\bibitem[Stockmeier et~al.(2018)Stockmeier, Kranert, Raming, Miller, Reimann,
  Rudolph, and Friedrich]{stockmeier2018edge}
L~Stockmeier, C~Kranert, G~Raming, A~Miller, C~Reimann, P~Rudolph, and
  J~Friedrich.
\newblock Edge facet dynamics during the growth of heavily doped n-type silicon
  by the czochralski-method.
\newblock \emph{Journal of Crystal Growth}, 491:\penalty0 57--65, 2018.

\bibitem[Stockmeier et~al.(2019)Stockmeier, Kranert, Fischer, Epelbaum,
  Reimann, Friedrich, Raming, and Miller]{stockmeier2019analysis}
L~Stockmeier, C~Kranert, P~Fischer, B~Epelbaum, C~Reimann, J~Friedrich,
  G~Raming, and A~Miller.
\newblock Analysis of the geometry of the growth ridges and correlation to the
  thermal gradient during growth of silicon crystals by the czochralski-method.
\newblock \emph{Journal of Crystal Growth}, 515:\penalty0 26--31, 2019.

\bibitem[Kellerman et~al.(2016)Kellerman, Kernan, Helenbrook, Sun, Sinclair,
  and Carlson]{kellerman2016floating}
Peter Kellerman, Brian Kernan, Brian~T Helenbrook, Dawei Sun, Frank Sinclair,
  and Frederick Carlson.
\newblock Floating silicon method single crystal ribbon--observations and
  proposed limit cycle theory.
\newblock \emph{Journal of Crystal Growth}, 451:\penalty0 174--180, 2016.

\bibitem[Fujiwara(2012)]{fujiwara2012crystal}
Kozo Fujiwara.
\newblock Crystal growth behaviors of silicon during melt growth processes.
\newblock \emph{International Journal of Photoenergy}, 2012, 2012.

\bibitem[Lau~Jr et~al.(2020)Lau~Jr, Maeda, Fujiwara, and Lan]{lau2020situ}
Victor Lau~Jr, Kensaku Maeda, Kozo Fujiwara, and Chung-wen Lan.
\newblock In situ observation of the solidification interface and grain
  boundary development of two silicon seeds with simultaneous measurement of
  temperature profile and undercooling.
\newblock \emph{Journal of Crystal Growth}, 532:\penalty0 125428, 2020.

\bibitem[Hu et~al.(2020)Hu, Maeda, Shiga, Morito, and Fujiwara]{hu2020situ}
Kuan-Kan Hu, Kensaku Maeda, Keiji Shiga, Haruhiko Morito, and Kozo Fujiwara.
\newblock In situ observation of multiple parallel (1 1 1) twin boundary
  formation from step-like grain boundary during si solidification.
\newblock \emph{Applied Physics Express}, 13\penalty0 (10):\penalty0 105501,
  2020.

\bibitem[Segal et~al.(1998)Segal, Vuik, and Vermolen]{segal1998conserving}
Guus Segal, Kees Vuik, and Fred Vermolen.
\newblock A conserving discretization for the free boundary in a
  two-dimensional stefan problem.
\newblock \emph{Journal of Computational Physics}, 141\penalty0 (1):\penalty0
  1--21, 1998.

\bibitem[Vuik et~al.(2000)Vuik, Segal, and Vermolen]{vuik2000conserving}
Cornelis Vuik, A~Segal, and Fred~J Vermolen.
\newblock A conserving discretization for a stefan problem with an interface
  reaction at the free boundary.
\newblock \emph{Computing and visualization in science}, 3\penalty0
  (1-2):\penalty0 109--114, 2000.

\bibitem[King et~al.(1999)King, Riley, and Wallman]{king1999two}
JR~King, DS~Riley, and AM~Wallman.
\newblock Two--dimensional solidification in a corner.
\newblock \emph{Proceedings of the Royal Society of London. Series A:
  Mathematical, Physical and Engineering Sciences}, 455\penalty0
  (1989):\penalty0 3449--3470, 1999.

\bibitem[Yeckel(2017)]{yekel2017}
Andrew Yeckel.
\newblock Novel means to modify growth interface shape in vertical bridgman
  crystal growth.
\newblock \url{http://www.cats2d.com/SIVB/insulatedVB.html}, 2017.
\newblock Accessed: 2021-01-31.

\bibitem[Kuiken(1988)]{kuiken1988note}
HK~Kuiken.
\newblock A note on the wall singularity of a solid--liquid interface caused by
  a difference between the thermal conductivities of the solid and the liquid
  phases.
\newblock \emph{SIAM Journal on Applied Mathematics}, 48\penalty0 (4):\penalty0
  921--924, 1988.

\bibitem[Ostrogorsky et~al.(2018)Ostrogorsky, Riabov, and
  Dropka]{ostrogorsky2018interface}
AG~Ostrogorsky, V~Riabov, and N~Dropka.
\newblock Interface control by rotating submerged heater/baffle in vertical
  bridgman configuration.
\newblock \emph{Journal of Crystal Growth}, 498:\penalty0 269--276, 2018.

\bibitem[Volz et~al.(2009)Volz, Mazuruk, Aggarwal, and
  Cr{\"o}ll]{volz2009interface}
MP~Volz, K~Mazuruk, MD~Aggarwal, and A~Cr{\"o}ll.
\newblock Interface shape control using localized heating during bridgman
  growth.
\newblock \emph{Journal of crystal growth}, 311\penalty0 (8):\penalty0
  2321--2326, 2009.

\bibitem[Sackinger et~al.(1989)Sackinger, Brown, and
  Derby]{sackinger1989finite}
PA~Sackinger, RA~Brown, and JJ~Derby.
\newblock A finite element method for analysis of fluid flow, heat transfer and
  free interfaces in czochralski crystal growth.
\newblock \emph{International journal for numerical methods in fluids},
  9\penalty0 (4):\penalty0 453--492, 1989.

\bibitem[Ramachandran and Gunjal(2009)]{ramachandran2009comparison}
PA~Ramachandran and PR~Gunjal.
\newblock Comparison of boundary collocation methods for singular and
  non-singular axisymmetric heat transfer problems.
\newblock \emph{Engineering analysis with boundary elements}, 33\penalty0
  (5):\penalty0 704--716, 2009.

\bibitem[Anderson and Davis(1994)]{anderson1994fluid}
DM~Anderson and SH~Davis.
\newblock Fluid flow, heat transfer, and solidification near tri-junctions.
\newblock \emph{Journal of crystal growth}, 142\penalty0 (1-2):\penalty0
  245--252, 1994.

\bibitem[Wigley(1969)]{wigley1969method}
Neil~M Wigley.
\newblock On a method to subtract off a singularity at a corner for the
  dirichlet or neumann problem.
\newblock \emph{Mathematics of Computation}, 23\penalty0 (106):\penalty0
  395--401, 1969.

\bibitem[Cheng et~al.(2014)Cheng, Han, Yao, Niu, and Recho]{cheng2014analysis}
Changzheng Cheng, Zhilin Han, Shanlong Yao, Zhongrong Niu, and Naman Recho.
\newblock Analysis of heat flux singularity at 2d notch tip by singularity
  analysis method combined with boundary element technique.
\newblock \emph{Engineering Analysis with Boundary Elements}, 46:\penalty0
  1--9, 2014.

\bibitem[Glicksman and Voorhees(1983)]{glicksman1983analysis}
ME~Glicksman and PW~Voorhees.
\newblock Analysis of morphologically stable horizontal ribbon crystal growth.
\newblock \emph{Journal of Electronic Materials}, 12\penalty0 (1):\penalty0
  161--179, 1983.

\bibitem[Helenbrook(2015)]{helenbrook2015solidification}
Brian~T Helenbrook.
\newblock Solidification along a wall or free surface with heat removal.
\newblock \emph{Journal of Crystal Growth}, 418:\penalty0 79--85, 2015.

\bibitem[Pirnia and Helenbrook(2020)]{pirnia2020analysis}
Alireza Pirnia and Brian~T Helenbrook.
\newblock Analysis of faceted solidification in the horizontal ribbon growth
  crystallization process.
\newblock \emph{Journal of Crystal Growth}, page 125958, 2020.

\bibitem[Oliveros et~al.(2015)Oliveros, Sridhar, and
  Ydstie]{oliveros2015existence}
German~A Oliveros, Seetharaman Sridhar, and B~Erik Ydstie.
\newblock Existence and static stability of the meniscus in horizontal ribbon
  growth.
\newblock \emph{Journal of Crystal Growth}, 411:\penalty0 96--105, 2015.

\bibitem[Kellerman(2013)]{kellerman2013floating}
Peter Kellerman.
\newblock Floating silicon method.
\newblock Technical report, Applied Materials-Varian Semiconductor Equipment,
  2013.

\bibitem[Zoutendyk(1978)]{zoutendyk1978theoretical}
John~A Zoutendyk.
\newblock Theoretical analysis of heat flow in horizontal ribbon growth from a
  melt.
\newblock \emph{Journal of Applied Physics}, 49\penalty0 (7):\penalty0
  3927--3932, 1978.

\bibitem[Zoutendyk(1980)]{zoutendyk1980analysis}
John~A Zoutendyk.
\newblock Analysis of forced convection heat flow effects in horizontal ribbon
  growth from the melt.
\newblock \emph{Journal of Crystal Growth}, 50\penalty0 (1):\penalty0 83--93,
  1980.

\bibitem[Daggolu et~al.(2012)Daggolu, Yeckel, Bleil, and
  Derby]{daggolu2012thermal}
Parthiv Daggolu, Andrew Yeckel, Carl~E Bleil, and Jeffrey~J Derby.
\newblock Thermal-capillary analysis of the horizontal ribbon growth of silicon
  crystals.
\newblock \emph{Journal of crystal growth}, 355\penalty0 (1):\penalty0
  129--139, 2012.

\bibitem[Daggolu et~al.(2013)Daggolu, Yeckel, Bleil, and
  Derby]{daggolu2013stability}
Parthiv Daggolu, Andrew Yeckel, Carl~E Bleil, and Jeffrey~J Derby.
\newblock Stability limits for the horizontal ribbon growth of silicon
  crystals.
\newblock \emph{Journal of crystal growth}, 363:\penalty0 132--140, 2013.

\bibitem[Daggolu et~al.(2014)Daggolu, Yeckel, and Derby]{daggolu2014analysis}
Parthiv Daggolu, Andrew Yeckel, and Jeffrey~J Derby.
\newblock An analysis of segregation during horizontal ribbon growth of
  silicon.
\newblock \emph{Journal of crystal growth}, 390:\penalty0 80--87, 2014.

\bibitem[Helenbrook et~al.(2016)Helenbrook, Kellerman, Carlson, Desai, and
  Sun]{helenbrook2016experimental}
Brian~T Helenbrook, Peter Kellerman, Frederick Carlson, Nandish Desai, and
  Dawei Sun.
\newblock Experimental and numerical investigation of the horizontal ribbon
  growth process.
\newblock \emph{Journal of Crystal Growth}, 453:\penalty0 163--172, 2016.

\bibitem[Greenlee(2015)]{greenlee2015towards}
Alison~S Greenlee.
\newblock \emph{Towards the development of a horizontal ribbon growth process
  to produce thin, monocrystalline silicon sheets via the stabilization of the
  (111) plane in undercooled melts}.
\newblock PhD thesis, Massachusetts Institute of Technology, 2015.

\bibitem[Sun et~al.(2018)Sun, Ding, Jiang, Xu, and Yuan]{sun2018simulating}
Tao Sun, Jianning Ding, Cunhua Jiang, Jiawei Xu, and Ningyi Yuan.
\newblock Simulating the horizontal growth process of silicon ribbon.
\newblock \emph{AIP Advances}, 8\penalty0 (8):\penalty0 085307, 2018.

\bibitem[Sun et~al.(2020)Sun, Zhang, Cheng, Zhu, Xu, Yuan, and
  Ding]{sun2020numerical}
Tao Sun, Zhongqiang Zhang, Guanggui Cheng, Keqian Zhu, Jiawei Xu, Ningyi Yuan,
  and Jianning Ding.
\newblock Numerical investigation of thermocapillary and buoyancy convection in
  horizontal ribbon growth with lid-driven boundary.
\newblock \emph{AIP Advances}, 10\penalty0 (11):\penalty0 115310, 2020.

\bibitem[Ke et~al.(2017)Ke, Khair, and Ydstie]{ke2017effects}
Jiaying Ke, Aditya~S Khair, and B~Erik Ydstie.
\newblock The effects of impurity on the stability of horizontal ribbon growth.
\newblock \emph{Journal of Crystal Growth}, 480:\penalty0 34--42, 2017.

\bibitem[Noronha et~al.(2020)Noronha, Oliveros, and
  Ydstie]{noronha2020weierstrass}
Eyan~P Noronha, German~A Oliveros, and B~Erik Ydstie.
\newblock Weierstrass’ variational theory for analysing meniscus stability in
  ribbon growth processes.
\newblock \emph{Journal of Crystal Growth}, 553:\penalty0 125945, 2020.

\bibitem[Lax and Wendroff(1960)]{lax1960systems}
Peter Lax and Burton Wendroff.
\newblock Systems of conservation laws.
\newblock \emph{Communications on Pure and Applied Mathematics}, 13\penalty0
  (2):\penalty0 217--237, 1960.

\bibitem[McDonough(2008)]{mcdonough2008lectures}
James~M McDonough.
\newblock Lectures on computational numerical analysis of partial differential
  equations, 2008.

\bibitem[Kudo(1980)]{kudo1980improvements}
Bosshi Kudo.
\newblock Improvements in the horizontal ribbon growth technique for single
  crystal silicon.
\newblock \emph{Journal of Crystal Growth}, 50\penalty0 (1):\penalty0 247--259,
  1980.

\bibitem[Cheng(1993)]{cheng1993growth}
Yan Cheng.
\newblock The growth of dislocation-free bicrystal silicon ribbon.
\newblock \emph{Journal of Physics D: Applied Physics}, 26\penalty0
  (7):\penalty0 1109, 1993.

\bibitem[Jewett(1981)]{jewett1981method}
David~N Jewett.
\newblock Method and apparatus for producing crystalline ribbons, September~15
  1981.
\newblock US Patent 4,289,571.

\bibitem[Howel et~al.(2011)Howel, Siegel, and Mengu]{howel2011thermal}
John~R Howel, Robert Siegel, and MP~Mengu.
\newblock Thermal radiation heat transfer 5th edition, 460-466, 2011.
\newblock URL \url{http://www.thermalradiation.net/sectionb/B-71.html}.

\end{thebibliography}
